# Warp: a method for neural network interpretability applied to gene expression profiles

Assya Trofimov[*123]     Sebastien Lemieux[12]     Claude Perreault[13]

[1-]Institute for Research in Immunology and Cancer (IRIC), [2-]Department of Computer Science and Research Operations department and [3-]Department of Medecine, University of Montreal, Quebec, Canada

**Abstract**

*We show a proof of principle for **warping**, a method to interpret the inner working of neural networks in the context of gene expression analysis. Warping is an efficient way to gain insight to the inner workings of neural nets and make them more interpretable. We demonstrate the ability of warping to recover meaningful information for a given class on a sample-specific individual basis. We found warping works well in both linearly and nonlinearly separable datasets. These encouraging results show that warping has a potential to be the answer to neural networks interpretability in computational biology.*

## 1 Introduction

Application of machine learning is off to a promising start in the field of computational biology. Every year, innovations in the field help bridge a little more the gap between measured biological data and observed phenotype. However, in some cases, algorithms are applied in a "black box" setting, which hinders the proper use of learning algorithms. In an image classification setting, it is expected that, while learning, the neural nets (NN) extract the representation of visual characteristics (shape, color) for a given class[1]. Notably, Simonyan and colleagues[2] show that when giving a pre-trained and frozen CNN an input of noise, they can adjust the input to maximize the class score and this way generate what seems to be the model's captured class notion. Using this type of analysis allows to visualize what the CNN detects[3] and what it learned about the characteristics of a given class. Moreover, recent work from Mikolov and colleagues[4] have shown through vector arithmetic in the learned representation space that learning algorithms extract meaningful data representations.

Unlike the field of image analysis, however, machine learning in computational biology calls for a more challenging problem. Image analysis is about training a machine to recognize images, a task at which humans excel. In contrast, machine learning algorithms in computational biology are asked to do the humanly impossible: crunch tremendous amounts of data of abstract construct and extract patterns that can be exploited to predict a phenotype. Some algorithms, such as random forests, are more readily interpretable, which is probably the reason they are so popular in biological data analysis. Indeed, random forests offer easy access to variables and thresholds deemed important for the prediction. In contrast, NN offer a bigger challenge in terms of interpretability and are often overlooked as good candidates for prediction tasks, since users want to avoid a "black box" situation and prefer to fully understand how the analysis is done.

In the following study, we propose a proof-of-principle to the difficult interpretability of NN in the analysis of biological data. For simplicity, we limit ourselves to the specific task of phenotype prediction based on gene expression data. However, we think that our solution has a broader applicability for tasks of different nature. Inspired by the work of Simonyan(*2*) and Mikolov(*4*), our method calculates the change necessary in the input dimension space, to classify the sample as one of the other class, while remaining on the data manifold. Here we define the technique and examine its behavior on a real gene expression dataset.

## 2   Methods

### 2.1 Datasets

Two datasets were selected for the experiments in this study. The first dataset called *XOR* is a standard visualization and testing dataset that is custom generated. It consists of 2-dimensional data points from two classes (here N=100), arranged in a non-linear fashion. The second dataset, *Phenotype*, was obtained from The Cancer Genome Atlas (TCGA), a National Institute of Health (NIH) cancer research consortium. We used the normal lung and kidney control RNA-Seq of 100 patients.

### 2.3 Input dimension interpolation

We pre-trained a multi-layer perceptron to classify the *XOR* dataset. Once the trained neural net architecture in place, we froze the layers, making updates to weights and biases impossible. We then added an additional layer of the same size as the input between the input and the first layer and only allowed updates to biases. We then proceeded to create a "new" dataset, where all members of a specific class were flipped to the other one. The network was then presented with one example and allowed to update the added bias layer and reach convergence. The bias layer (regularized by L2 to promote the selection of small warp vector values) was then extracted and the values were examined to determine the smallest necessary adjustment in input dimension space to classify the sample as one of the other class.

### 2.4 Variational Autoencoder and latent space vector arithmetic

Our variational autoencoder (VAE) (*5*) encodes gene expression data into smaller representation space, while maximizing the likelihood of the data *X* in representation space *Z* (Gaussian Z).

$$p(x) \approx \frac{1}{n} \sum_i p(X|z_i)$$

The general cost function for the VAE is:

$$\mathcal{L}(\zeta, \theta | X) = -D_{KL}(q_\zeta(z|X) || p_\theta(z)) + \mathbb{E}_{q_\zeta(z|X)}(log(p_\theta(X|z)))$$

Briefly, for parameters $\varsigma$ and $\theta$, the network maximizes the likelihood of finding a good representation in *Z* space for every *X,* while minimizing the reconstruction error when decoding *z* to *X*. We performed latent space vector arithmetic, similar to Mikolov(*4*), by using the general class centroids and show that the VAE has extracted meaningful representations.

# 3 Results

## 3.1 Intuition on synthetic data

We define the term *class gene signature* as a specific set of genes as well as their expression levels, that are sufficient to identify a phenotype or class. The standard approach to examining class gene signatures would be to calculate for each class, a mean gene expression profile and then examine the difference. This technique does not capture class-specific patterns that are not linearly separable. Our first experiment was performed on the *XOR* dataset (Figure 1A). A standard class mean profile would not work in this case, since the circles are concentric. We then implemented and trained an MLP on the dataset and proceeded to perform the warp for all examples of class A (red data points).

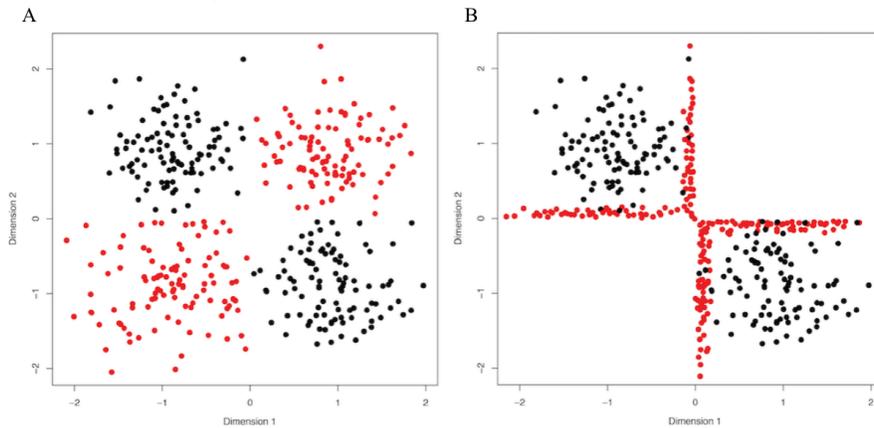

Figure 1 – XOR dataset linear interpolation method.

When applying the warping vectors to class A, we observe that all data points seem to have migrated inside of the decision boundaries (Figure 1B). This confirms that our approach functions well with non-linearly separable datasets and highlights the complexity of the output, since bias vectors are calculated on a sample-specific level.

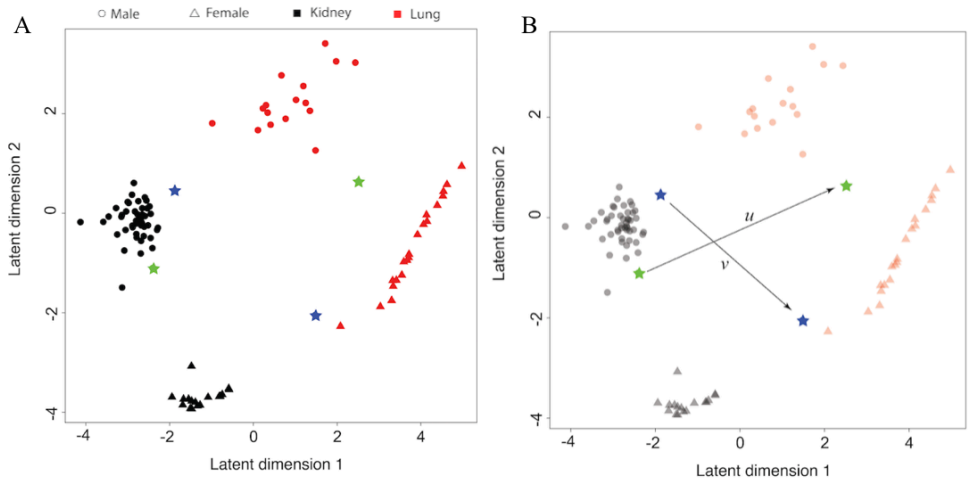

Figure 2 A) Plot of examples in representation space post training. B) Representation space group centroid vector arithmetic.

## 3.2 Important flaw in the linear interpolation model

This linear interpolation technique is however flawed, since it assumes the data manifold is continuous in input space. Since this assumption is likely false, this technique will necessarily create adversarial examples (*6*). Since the representation learning task is to maximize efficient representation of the data manifold within the representation space, we propose to move the interpolation to the hidden layers, thus guaranteeing that any new point in the space will belong to a plausible example.

### 3.3 Corrected proof of concept with sex and tissue source-site classification

A VAE is used to encode the *Phenotype* dataset. We observe that the plotted coordinates in representation space seem to cluster by group (Figure 2A). We then performed interpolation in representation space similar to the technique described in Section 3.1 (Figure 2B). For an example point with the class labels *male* and *lung*, the coordinates in representation space are [0.672, 2.77]. When added the female centroid and kidney centroid and then subtracted the male and lung centroids, we obtain a new point coordinate, [-1.147, -2.46] (Figure 3).

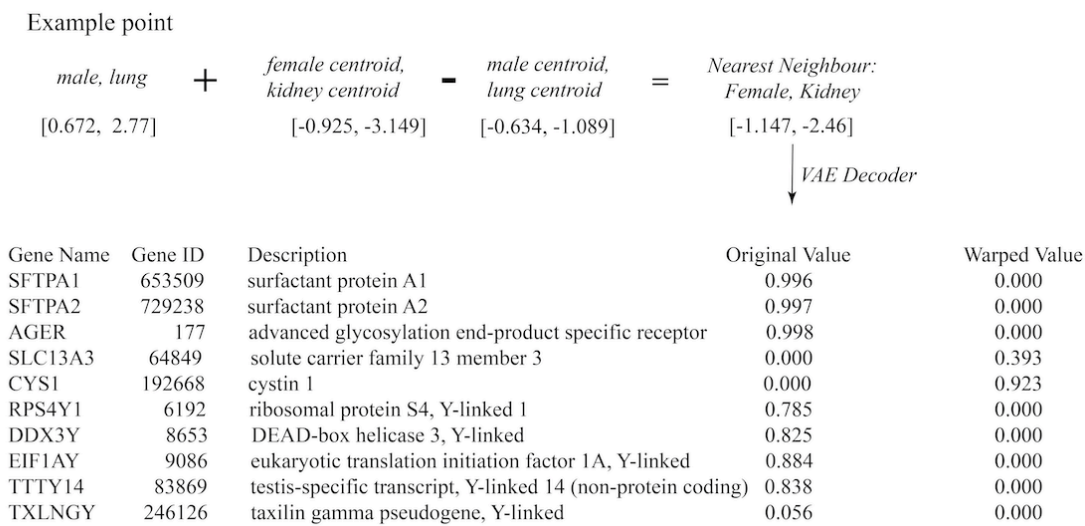

Figure 3- Example of vector arithmetic in representation space for class information

The nearest neighboring points hold the labels *female* and *kidney*. Moreover, the decoder part of the VAE allows the decoding of this new point into input dimensions, yielding the exact adjustment values for that example point to become one of the other class (top 10 genes to adjust are shown) (Figure 3). Indeed, originally, the patient expressed genes of the sufractant family(*7*), as well as chromosome Y-linked genes. In order to be classified as *female* and *kidney*, the network has suggested that these genes be adjusted to an expression value of 0 and simultaneously boost the expression value of the solute carrier family proteins as well as the cystin family, two genes known to be overexpressed in kidney cells(*8*)(*9*).

## 5  Conclusion

In this proof-of-principle, our method shows encouraging results as a valid alternative for examining class gene signatures post neural network training. The approach offers great flexibility, detecting various types of signatures. Ongoing experiments are being conducted to determine how destructive a warp is to other gene signatures within the samples as well as overcoming reconstruction blurriness caused by the VAE nature of encoding.